# Using Markov Decision Process to Model Deception for Robotic and Interactive Game Applications

Ali Ayub[1], Aldo Morales[2], and Amit Banerjee[3]

*Abstract*— **This paper investigates deception in the context of motion using a simulated mobile robot. We analyze some previously designed deceptive strategies on a mobile robot simulator. We then present a novel approach to adaptively choose target-oriented deceptive trajectories to deceive humans for multiple interactions. Additionally, we propose a new metric to evaluate deception on data collected from the users when interacting with the mobile robot simulator. We performed a user study to test our proposed adaptive deceptive algorithm, which shows that our algorithm deceives humans even for multiple interactions and it is more effective than random choice of deceptive strategies.**

## I. Introduction

Current research on computer games has been focused on improvement of search techniques [1], artificial intelligence [2] and imparting effective information to the user [3]. For the latter, there is a natural counterpart: deception. Deception imparts wrong information or just conceals it completely and has a long history related to the study of intelligent systems. According to biologists and psychologists, deception provides an evolutionary advantage for the deceiver [4]. It has also been noted that higher-level primates use deception, which serves as an indicator of the theory of mind [5]. Animals use different types of deception mechanisms to survive. Chimpanzees, for example, deceive based on the situation [6]. They can deceive an animal or human depending on their objectives.

Robots in intelligent systems can also gain an advantage over rivals by practicing deceptive behavior. For example, one application where robot deception has an impact is in the military [7] and real time strategy games. Another recent work studied and modeled deception in hardware trojan detection [8]. Although deception has plenty of potential benefits, there has been limited work done on deception in robots or interactive games. Further, most of the approaches are designed to deceive humans only for a single or a small set of interactions. No effort has been made in the past to develop an algorithm that can deceive humans for multiple interactions.

This paper investigates the use of deceptive strategies proposed in [9] on an autonomous mobile robot simulator that can be used in interactive computer games. Particularly, the purpose of this research is to develop an algorithm, for a mobile robot simulator, that can deceive humans for multiple interactions i.e. even if the humans have seen all the possible trajectories, the algorithm should still be able to deceive them. We believe that introducing deception in interactive computer games in the consumer electronics field can enhance the user experience and entertainment value of these games [10].

This paper has three main objectives: 1) Extend the deceptive strategies in [9] and implement them on an autonomous mobile robot simulator 2) Develop a general algorithm to use these deceptive trajectories to deceive humans for multiple interactions, 3) Develop a mathematical evaluation model to determine if the humans are indeed deceived by the robot. We show through a user study that humans can be deceived by a simulated mobile robot for multiple interactions using the proposed algorithm and our approach is better at deceiving humans than just a simple random selection of deceptive strategies. Further, most users stated that they were entertained by the deceptive robot, especially in the later interactions.

## II. Related Work

There has been limited amount of research done in deception using mobile robots via motion in recent years. For example, Dragan et al. [9] developed different deceptive trajectories (termed: exaggerating, switching and ambiguous) in the case of a two target system and studied those trajectories' deceptive effects on human participants. They used a 2 Degree of Freedom (DOF) robotic arm to generate the deceptive trajectories and performed different user studies. Their results showed that the three strategies are deceptive when humans interact with them once, but they are ineffective when the humans interact with them multiple times. To deceive humans for multiple interactions, they developed six different trajectories using the combination of their initial strategies, but their experiments were limited. For example, they only used the six different trajectories for six iterations. All these six trajectories were fixed, and the robot did not choose them in real time. Hence, this experiment was similar to interacting with six different trajectories once and if a human interacts with the robot again, he/she will not be deceived by these six trajectories. Moreover, in their study, users were asked to guess the target in the middle of each strategy. This does not provide good information on whether the users were deceived because there is a 50% chance of guessing correctly. In this paper, we present a simple GUI for the user study in which we collect the predictions of the users for the entire interactions.

Wagner et al. [11] developed a game-theory based deception approach using a mobile robot in a hide-and-seek scenario to deceive other mobile robots. The deceiver robot used the model of the robot being deceived, for deception. Although the approach was general for deceiving other robots with a known model, it cannot be used to deceive humans because of their ability to remember previous experiences.

[1]Department of Electrical Engineering at The Pennsylvania State University, State College, PA, 16802. (email: aja5755@psu.edu).

[2]Department of Electrical Engineering, Penn State Harrisburg, Middletown, PA 17057 USA (e-mail: awm2@psu.edu).

[3]Department of Mechanical Engineering, Penn State Harrisburg, Middletown, PA 17057 USA (e-mail: aub25@psu.edu).

In this paper, we treat deceiving humans for multiple interactions as a memory problem since humans can remember previous interactions with the robot. Markov chains have been used in the past to model memory-based systems [12], therefore they could also be used to model deception. Nonlinear Markov games have been used in the past to model deception in the context of control theory [13, 14] but the work was limited since no user studies were conducted. We propose a variation of the Markov decision process for choosing a deception strategy to deceive humans for multiple interactions.

## III. Deceptive Strategies On A Mobile Robot Simulator

The three deceptive trajectories developed in [9] were implemented on a robotic arm. We implemented the trajectories using the MATLAB robotics toolbox [15] on a mobile robot simulator. For the exaggerating trajectory, the robot was moved closer to the false target, using an optimal path, and then moved to the real target optimally. For the switching trajectory, the robot alternated between two targets horizontally while vertically moving towards the real target and for the ambiguous trajectory, the robot moved straight vertically at an equal distance from both targets and then moved towards the real target when it reached a certain vertical distance from both targets. Fig. 1 shows the three trajectories implemented on the mobile robot simulator:

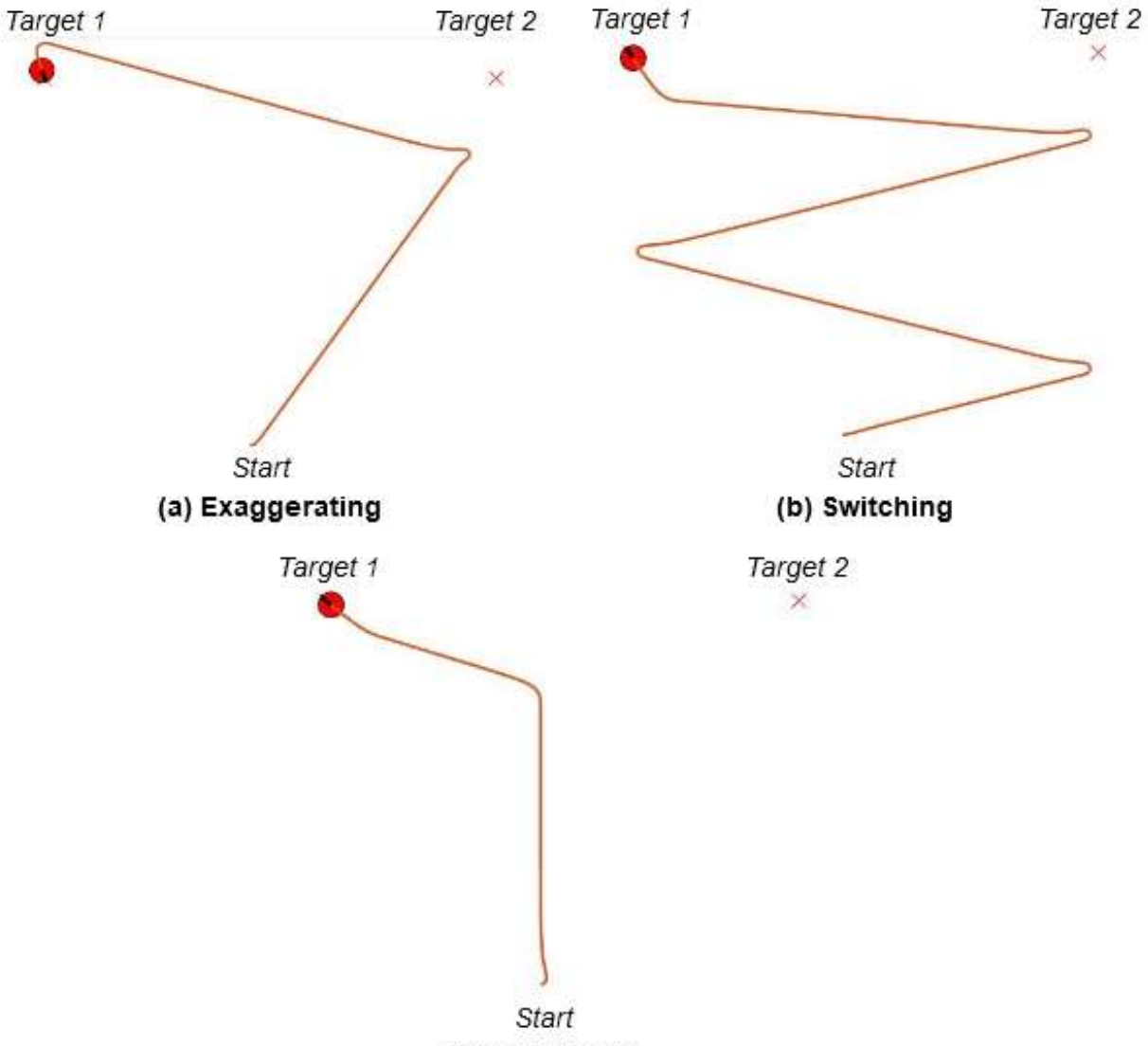

Fig. 1. Three main deceptive strategies on a mobile robot simulator

### *A. Version-2 of the main strategies*

Dragan et al. [9] conducted some surveys on how humans deceive when there are two targets available in an environment. Based on these surveys, some of the participants first used one of the three deceptive trajectories moving their hand towards a target but then back to the other target at the last moment. This observation was not used in [9] to deceive humans using the robotic arm. We hypothesize that this strategy can be helpful when deceiving humans for multiple interactions because once people see a deceptive trajectory, they will not get deceived again using the same trajectory.

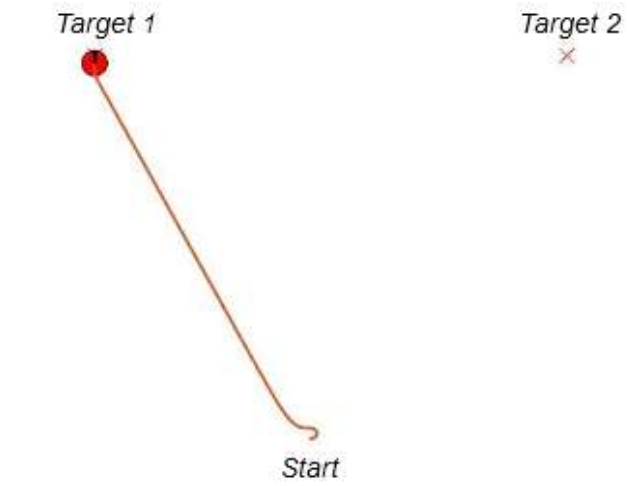

Fig. 2. Optimal strategy in which the robot simply just goes to the intended target

This variation can create some uncertainty for a person even if the trajectory looks familiar. In section VII, user study results show that the variation in these trajectories help maintain the deceptive effectiveness over multiple interactions.

To implement version-2 of the three trajectories, we made one addition to the main trajectories: move the robot back to the other target once it finishes the main trajectory. Fig. 3 shows version-2 of the three trajectories:

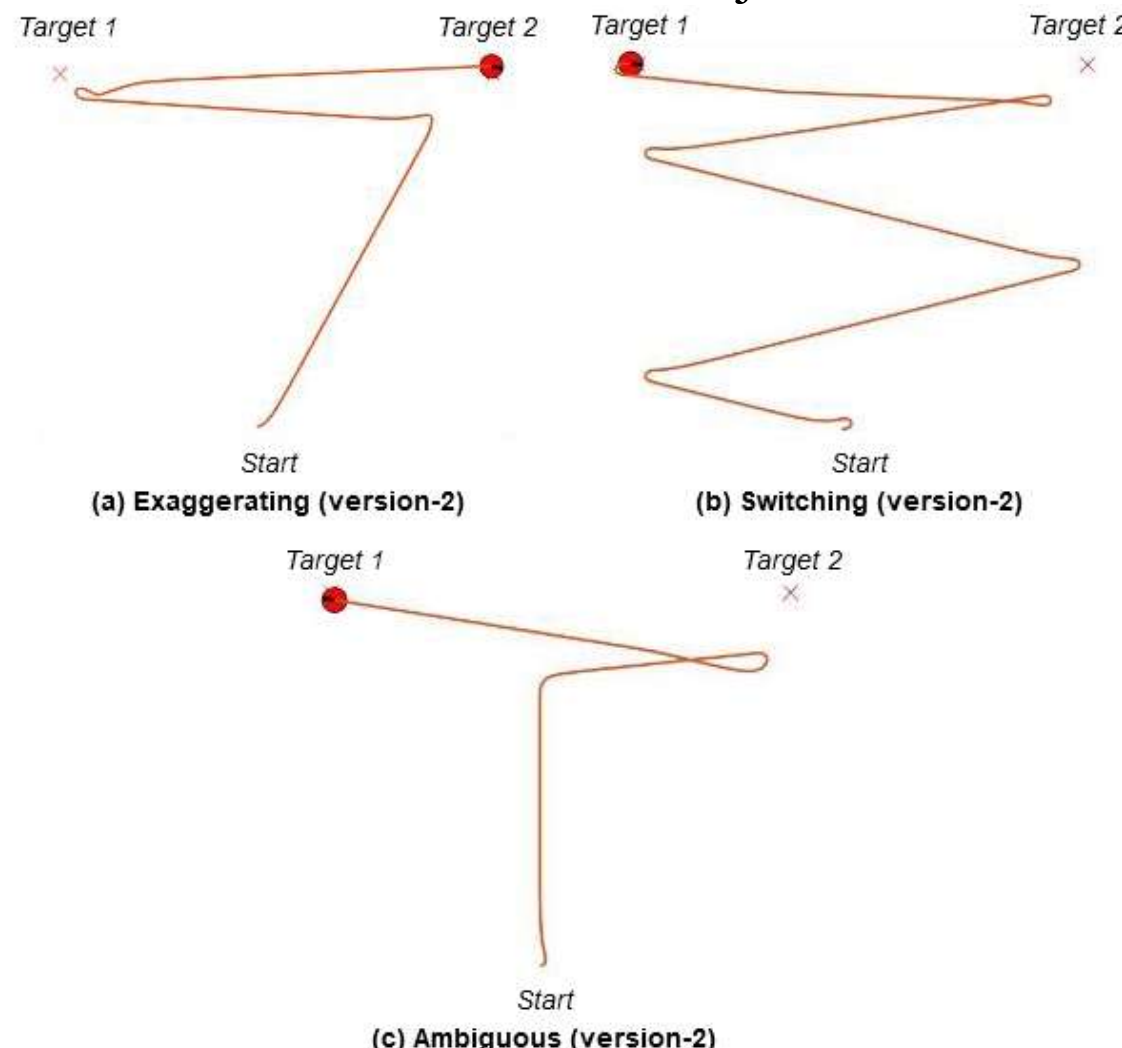

Fig. 3. Version 2 of the three deceptive strategies in which the robot goes back to the other target at the last moment

### ***B.*** *Optimal Strategy*

Another addition to the deception algorithm for multiple interactions is the inclusion of the optimal trajectory i.e. move the robot to the real target using an optimal path. This trajectory is not deceptive (it imparts true information) but we hypothesize that humans will get deceived by it once they have interacted with the robot and seen the other deceptive trajectories (particularly exaggerating because it is the same as the optimal trajectory in the start). The results shown in section VII prove our hypothesis. This trajectory (see Fig. 2) is implemented by optimally moving the robot to the real target.

## IV. Representing Deception For Multiple Interactions Using Markov Process

As mentioned earlier, even if a strategy is deceptive, once the humans interact with it for some iterations, they cannot be deceived again using the same strategy. Hence, we propose that the probability of selection of strategies in each interaction should be dependent upon the strategies chosen in previous interactions.

A simple Markov chain can be used to model the transition probabilities among different states (deceptive strategies) based on the choice of previous states. In this paper, we have the three main strategies and the optimal strategy to choose from at each interaction (we term a strategy as a state and an interaction as an iteration in the Markov process context). After the choice of the main strategy, there is a choice between the main version or version-2 of the strategies. Fig. 4 shows the first order Markov chain model of the four strategies with their corresponding transition probabilities. Selection between the main and version-2 can be similarly represented as two states with corresponding transition probabilities.

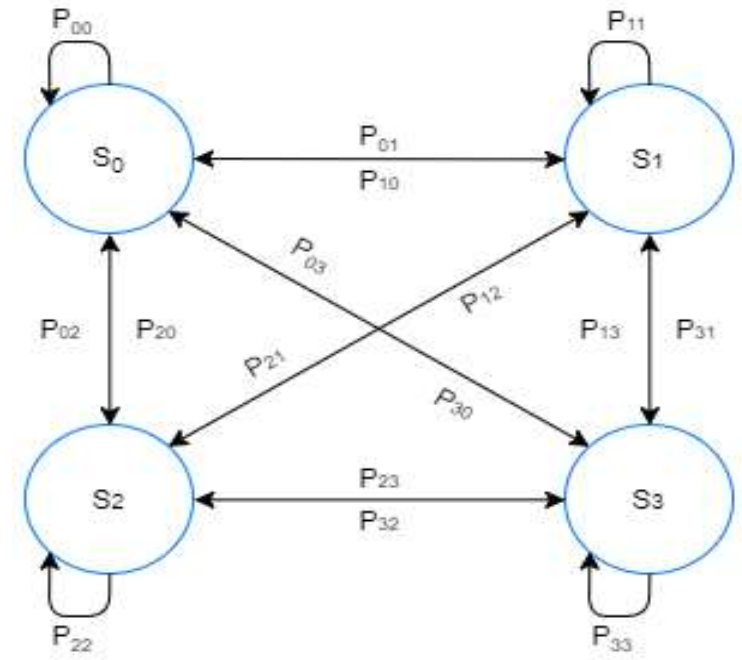

Fig. 4. Markov chain for four states with transition probabilities from each state to every other state.

Where S0, S1, S2, and S3 denote the exaggerating, switching, ambiguous and optimal trajectories, respectively and $P_{ij}$ $\forall i, j \in \{0,1,2,3\}$ represents the transition probability from state $i$ to state $j$. Since at each iteration the probability of occurrence of each strategy should be dependent upon the strategies selected in all the previous iterations, it could be modeled as a higher order Markov process. A higher order Markov process contains more history related to the previous states. The probability of occurrence of a state $S_i$ at iteration $j$ in an $n$th order Markov process can be modeled as:

$$P\left(S_{ij} \middle| S_{ij-1}, S_{ij-2}, \dots, S_{i1}\right) = P\left(S_{ij} \middle| S_{ij-1}, \dots, S_{ij-n}\right)$$

However, the higher order Markov process is not feasible for implementation because the order of the process will increase with the number of iterations. If the state-transition probabilities are not fixed, the memory required to keep track of the all the previous states and the corresponding transition probabilities will increase as well. To deal with this problem, we reset the probabilities of all the states to one iteration back, once they all have occurred at least once. Since all the states occur at least once, there is no need for the algorithm to remember this, hence all the probabilities are changed such that the states did not occur for one iteration.

## V. Deception Algorithms For Multiple Interactions

Generally, in a Markov process, the state transition probabilities are fixed and saved in a matrix. Using fixed probabilities for the strategies for all iterations can deceive humans for multiple iterations because of the random selection of the states. However, this approach is not optimal because all the transition probabilities are fixed at the same value (i.e. ¼ for four states for maximum uncertainty). Intuitively, due to the fixed probabilities, the chance of consecutively recurring strategies remains constant and the humans will not get deceived in such iterations because of the repetition. We call this approach the "random algorithm".

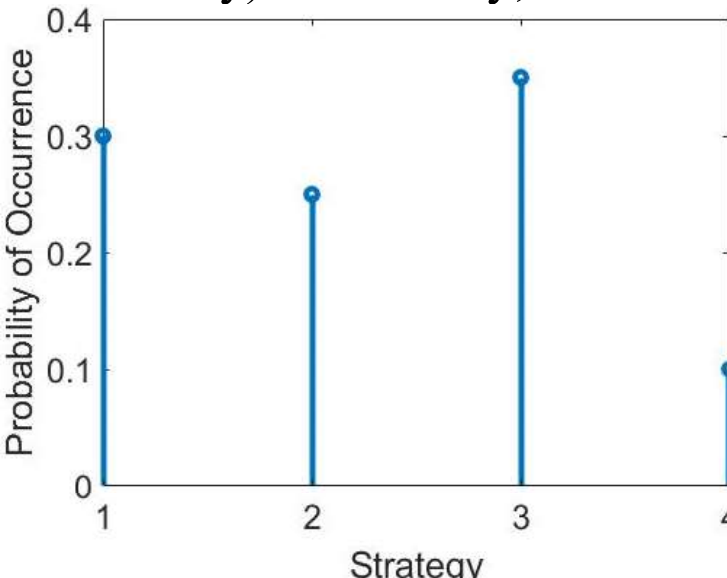

Fig. 5. Probabilities of occurrence for four states using the random algorithm in 100 iterations.

Entropy analysis of the random algorithm also provides some insight into why this approach is not perfect for deception. Entropy is the measure of uncertainty of a certain event and is defined as [16]:

$$H(S) = -\sum_{i=1}^{n} p(S_i) \log(p(S_i)) \quad (1)$$

Where $H(S)$ is the entropy of the random variable $S_i$ which represents the strategy choice i.e. 1, 2, 3 or 4 (exaggerating, switching, ambiguous or optimal), $n$ is the total number of states and $p(S_i)$ is the probability of occurrence of strategy $S_i$, in the total number of iterations (N), and is defined as:

$$p(S_i) = \frac{N_{S_i}}{N}$$

Where, $N_{S_i}$ is the number of times state $S_i$ occurs in $N$ iterations. If the experiment is run by randomly choosing a strategy at each iteration, the probability of selection of a state will be ¼ due to the uniform distribution. With no updates on this probability of selection of a state, there is an extremely low chance that all the states will occur an equal number of times within an experiment, which means that $p(S_i)$ will not be ¼. This reduces the entropy of the system, which in turn decreases deception. We conducted an experiment for 20 iterations (to model the experiment of multiple interactions with a human) and in each iteration one of the four states were chosen based on the uniform probability distribution. This experiment was run 10 times for robustness. We observed that the probability of occurrence of the states (averaged over the 10 runs of the experiment) was not uniform and the entropy was lower (1.79) than the maximum value of 2 (see Fig. 5) because some states occurred more times than the others with high repetition rate. These results show that the random algorithm is not the best for deceiving humans for multiple interactions. In section VII, results of the experiments to deceive human participants using the random algorithm affirm our hypothesis.

One way to ensure entropy remains maximum for multiple interactions is by forcing each strategy to occur equal number of times in all iterations by using a Markov decision process to remember previously occurred strategies. We call this strategy the “fixed algorithm”. This strategy maximizes entropy and seems to be perfect for deception for multiple interactions but fixing the total number of occurrences of states can have disadvantages of its own. There are two ways to implement the fixed algorithm. 1) Each state can be fixed to occur once in four iterations, 2) all states can be fixed to occur equal amount of times in the set of all iterations, while choosing each state randomly in a single iteration. Although, both these implementations of the fixed algorithm ensure maximum entropy, they are not optimal for deception for multiple interactions. Intuitively, for the first approach, since each state occurs once in a set of four iterations, a human can detect this pattern easily which will decrease deception. The second approach poses almost the same problem as the random algorithm in which a state can occur multiple times in a row or in a set of four iterations, hence decreasing deception.

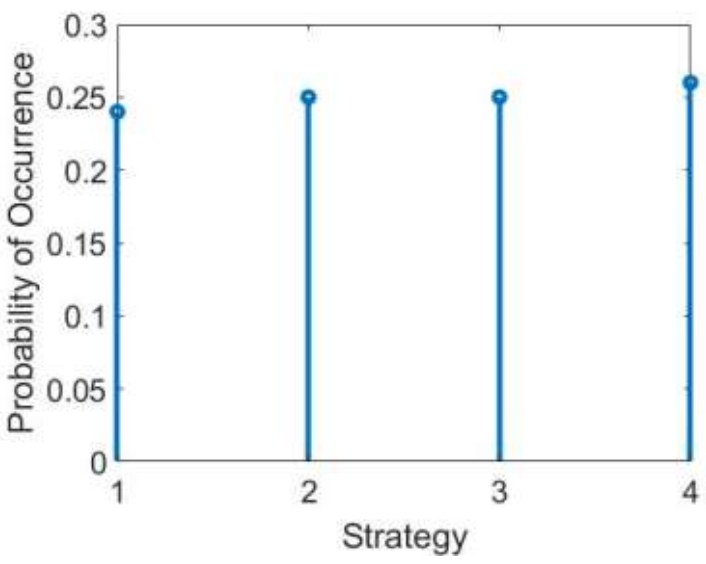

Fig. 7. Probabilities of occurrence for four states using the adaptive algorithm where lambda=0.5 in 100 iterations

To fix these issues, we developed an approach to transition a part of the probability of the states that occur more to the ones that occur less at a given iteration in the process. Formally, if a state $S_i$ occurs at iteration $k$, the probability of its occurrence $p(S_i)$ is calculated as:

$$p(S_i)_k = \lambda(1 - p(S_i)_{k-1}) \qquad (2)$$

Where, $p(S_i)_{k-1}$ is the probability of state $S_i$ at iteration $k-1$ and $\lambda$ is the transition rate parameter which controls how much the probability of occurrence of a state should be reduced when it occurs in an iteration. The part of the probability $\lambda p(S_i)_{k-1}$ is transitioned equally to all the other states' probabilities that have occurred less than the state $S_i$. In this way, our algorithm penalizes those states that occur less and prefers other states to be chosen that have occurred less in the long run of iterations. Intuitively, it can be observed from equation (2), that the higher the value of $\lambda$ the higher the amount of probability removed from a state after it occurs at an iteration. This increases the chance of other states to occur in the next iterations. If $\lambda = 0$, the approach will become the fixed algorithm and if it is 1 the approach will become the random algorithm. We ran a set of simulations at different values of $\lambda$ and 0.5 ensured a good tradeoff between entropy and repetition of states. We call this probability transition approach with $\lambda = 0.5$, the adaptive algorithm. Fig. 7 shows the experiment of 20 iterations using the adaptive algorithm with $\lambda = 0.5$. Similar to the random algorithm experiment, we performed this experiment 10 times for robustness. The entropy of the system (1.98) is not equal to the maximum value of 2, but it is only slightly lower and it is much higher than the entropy of the random algorithm and unlike the fixed algorithms the pattern is not easy to predict and has randomness in the selection of states. In section VII, results shown for experiments run with human participants using the adaptive algorithm affirm our analysis.

### *A. GUI For Data Collection*

Both the adaptive and the random algorithms were implemented in MATLAB and a simple GUI (Fig. 8) was developed for collecting deception data from humans. For every iteration, after the start button is pressed, the GUI shows the simulated robot moving towards one of the two targets using one of the deception strategies. The deception strategy can be selected by either the random or the adaptive algorithm. During the robot movement, the human observer can predict the robot's destination by moving the scrolling pad, at top of the GUI. The scrolling pad values range from 0 (left target) to 1 (right target) and for each interaction all the pad values are collected to analyze whether or how much the human was deceived by the corresponding deceptive strategy. In section VII we show the results of the data collected from humans through this GUI using the adaptive and random algorithms.

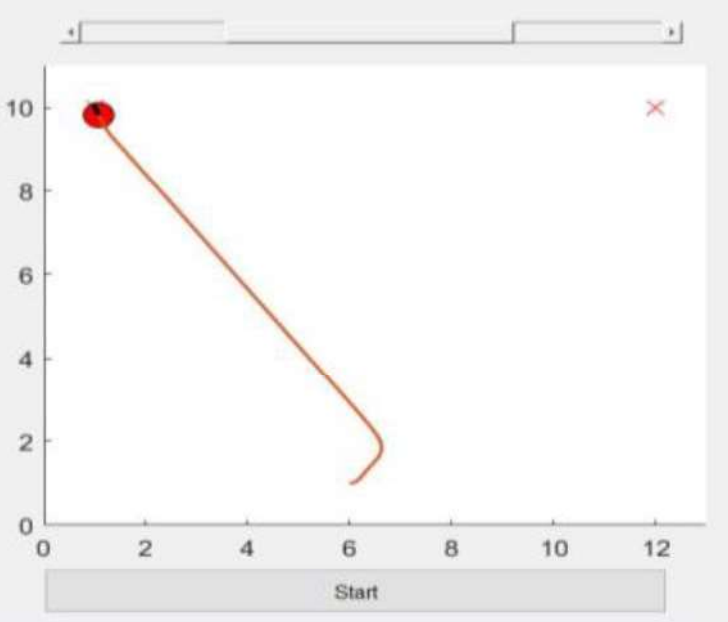

Fig. 8. GUI used for collecting human deception data. In the figure above, the robot uses the optimal strategy to move towards the left target.

## VI. EVALUATION METRIC TO RATE DECEPTION

Deception is defined as imparting false information or hiding true information. Based on this definition, a human observer interacting with the GUI will be considered deceived when he/she believes that the false target is the true one or he/she is uncertain about the true target. Hence, distance of the pad from the true target during an interaction shows the inaccuracy of the human in predicting the true target and movements of the pad show the uncertainty of the human about his/her prediction.

Based on these intuitions, we define two metrics to evaluate deception in humans: error and confusion. Error measures the distance of the pad from true target in an interaction, defined as:

$$Error = \frac{1}{\tau}\int_0^{\tau} |T - \mu(t)| dt \qquad (3)$$

Where T represents the true target (0 or 1), $\mu(t)$ is the pad position at time $t$ and $\tau$ is the total time of an interaction with the robot. Error = 0 shows that the pad was at the true target for the entire time indicating no deception; similarly, Error = 1 indicates maximum deception.

Confusion measures the belief of the human in his/her choice of the target and is defined as:

$$Confusion = \frac{1}{\tau}\int_0^{\tau} t|\mu'(t)| dt \qquad (4)$$

Where, $\mu'(t)$ is the derivative of the pad position at time $t$. Since the pad can only be moved at a constant rate, $\mu'(t)$ will either be 0 (no movement) or a constant value (movement). A confusion value of 0 indicates that the user never moved the pad indicating 100% belief in his/her choice of the target (right or wrong) and confusion=1 (normalized) indicates maximum uncertainty. The rationale for putting *t* inside the integral is that as time goes on, the human acquires more data on the robot's behavior, so we expect motion at the end of the time interval to be more indicative of the human's confusion.

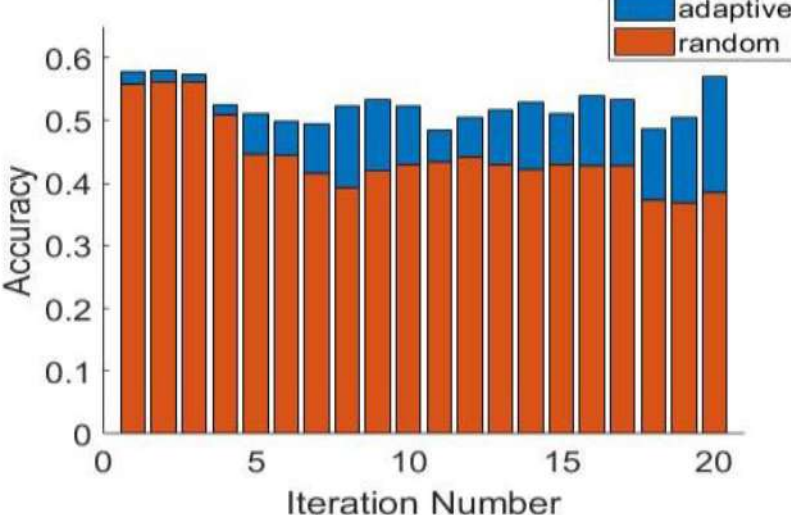

Fig. 9. Mean error values of the adaptive and random algorithms for 20 iterations

We exclude the data for the first 5% and the last 5% of the time of the interaction from the above-mentioned equations. Since targets are selected randomly and the robot starts at an equal distance from the target, the starting 5% of the time is just based on the guess of the user. During the last 5% of the time, the robot moves towards the true target; hence, collecting that data is pointless. Also, for the confusion equation (4), we exclude the initial pad movements toward a target and start counting after the participants start to move the pad in the opposite direction. The reason is because the pad starts in the middle, and the user has to move it towards one of the targets for prediction. In the next section, we present a user study to gather data from human participants over the course of 20 iterations using the adaptive and random algorithms and evaluate them using these metrics.

## VII. Experiments

We performed a human-computer interaction study for the evaluation of the deceptive strategies on a mobile robot simulator and the deceptive experiment for multiple interactions. In this study, we performed two experiments for the comparison of our proposed deception approach against the random choice of strategies for multiple interactions. A total of 60 participants (35 male, 25 female) were chosen between the ages of 18 to 25 from Penn State. 30 participants were randomly chosen to interact with the simulator with the proposed deception algorithm and other 30 with the random algorithm. The participants were shown a total of 20 iterations and were asked to move the scrolling pad on the top of the GUI to either of the targets using the left and right arrow keys on the keyboard in each iteration. The scrolling pad represented the target prediction of the participants. During each interaction, the pad positions of each participant were saved. Before the 20 iterations, participants played with the robot a couple of practice rounds in which the optimal strategy was shown to get familiar with the environment. We collected this data to compare the deceptive effectiveness of the two algorithms. We proposed the following hypothesis:

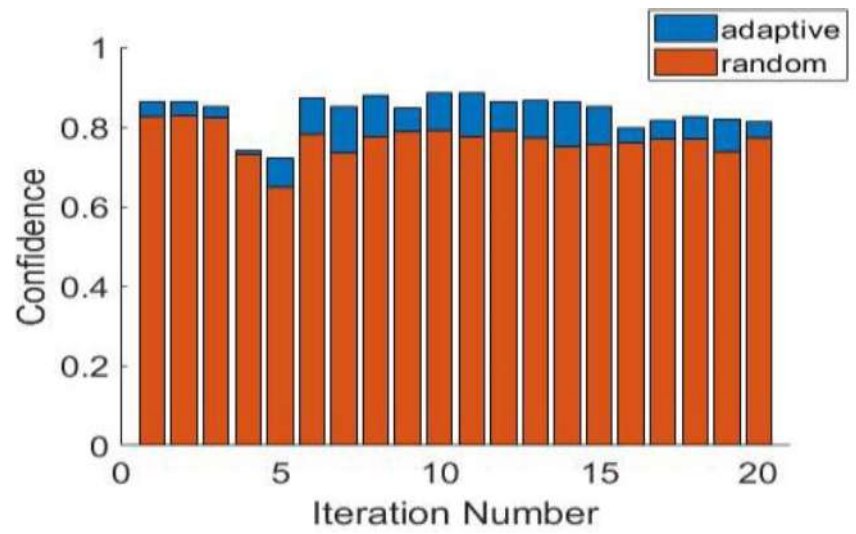

Fig. 10. Mean confusion values of the adaptive and random algorithms for 20 iterations

| | **Error (Reference Mean (R) = 0.5)** | | | **Confusion (Reference Mean (R) = 0.95)** | | |
|---|---|---|---|---|---|---|
| **Iteration Number** | Mean (M) | P Value | Inference | Mean (M) | P Value | Inference |
| **1** | 0.5703 | 0.0066 | M>R | 0.8504 | 0.2109 | M>R |
| **2** | 0.5403 | 0.0992 | M≈R | 0.8726 | 0.5241 | M≈R |
| **3** | 0.5112 | 0.7479 | M≈R | 0.7239 | 0.0201 | M>R |
| **4** | 0.5295 | 0.4494 | M≈R | 0.5426 | 1.5* $10^{-4}$ | M<R |
| **5** | 0.5333 | 0.1144 | M≈R | 0.8526 | 0.2759 | M≈R |
| **6** | 0.5108 | 0.5081 | M≈R | 0.8784 | 0.5189 | M≈R |
| **7** | 0.4991 | 0.9521 | M≈R | 0.8486 | 0.2484 | M≈R |
| **8** | 0.4936 | 0.6555 | M≈R | 0.8869 | 0.6562 | M≈R |
| **9** | 0.5225 | 0.1276 | M≈R | 0.8851 | 0.6181 | M≈R |
| **10** | 0.5333 | 0.0511 | M≈R | 0.8656 | 0.2665 | M≈R |
| **11** | 0.5233 | 0.1655 | M≈R | 0.8657 | 0.2861 | M≈R |
| **12** | 0.4848 | 0.3351 | M≈R | 0.8652 | 0.2787 | M≈R |
| **13** | 0.5040 | 0.7992 | M≈R | 0.8510 | 0.1366 | M≈R |
| **14** | 0.5174 | 0.4818 | M≈R | 0.7981 | 0.0565 | M≈R |
| **15** | 0.4775 | 0.3227 | M≈R | 0.8167 | 0.1064 | M≈R |
| **16** | 0.4802 | 0.3705 | M≈R | 0.8271 | 0.0942 | M≈R |
| **17** | 0.4729 | 0.2003 | M≈R | 0.8211 | 0.0856 | M≈R |
| **18** | 0.4747 | 0.2342 | M≈R | 0.8130 | 0.0632 | M≈R |
| **19** | 0.4873 | 0.5158 | M≈R | 0.8640 | 0.2744 | M≈R |
| **20** | 0.5042 | 0.8338 | M≈R | 0.8635 | 0.3366 | M≈R |

Table 1. Single sample t-test analysis of error and confusion values for 20 iterations

H1: *The adaptive algorithm is better than the random algorithm in deceiving humans for multiple interactions.*

Two-sampled t-test analysis (with 5% significance level) between the error and confusion values of the two algorithms (Table 1) shows that the error and confusion values for the first four iterations of the adaptive and random algorithms are similar, but for the rest of the iterations, adaptive algorithm has significantly higher values than the random algorithm except confusion value for the last iteration. These results prove H1. Fig. 9 and 10 show the mean error and confusion values respectively, of the adaptive and random algorithm for 20 iterations. They show that in the starting iterations both algorithms have similar error and confusion values because all the strategies are chosen for the first time but in later iterations, these values drop for the random algorithm but remain in the same range for the adaptive algorithm, which agrees with our t-test analysis results.

For the adaptive algorithm, after 20 iterations all participants were also asked to rate on a Likert scale of 1-7 (1 lowest, 7 highest) how intelligence, trust, deception and entertainment of interacting with the robot simulator changed. We also asked the participants if the robot's movement was intentional and all of them said yes. We proposed the following hypothesis:

H2: *Ratings for entertainment, deception and intelligence increase and decrease for trust.*

Single-sample t-test analyses (with 5% significance level) on the Likert scale data for entertainment, deception, intelligence and trust with reference mean of 3.5 (Entertainment: mean=6.077, p-value=4.3$*10^{-30}$; Deception: mean=4.808, p-value= 6.44$*10^{-22}$; Intelligence: mean=6.805, p-value=5.83$*10^{-33}$, Trust: mean=2.249, p-value=1.56$*10^{-20}$) show that entertainment, deception and intelligence were significantly higher than the reference mean and trust rating was significantly lower, which proves H5. Fig. 11 shows the mean Likert scale ratings for all four variables, which agree with the t-test analyses.

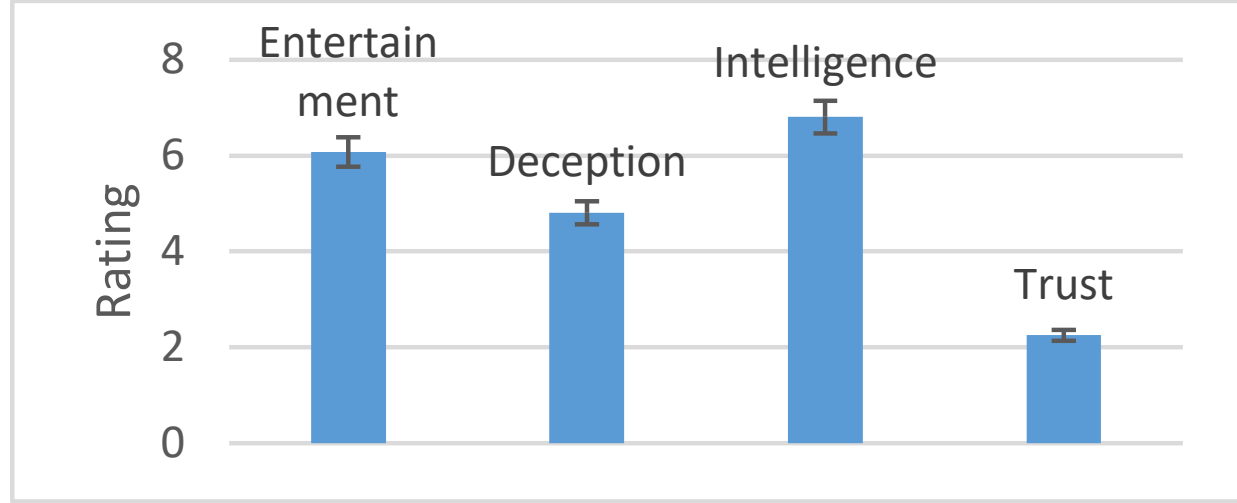

Fig. 11. Entertainment, deception, intelligence and trust ratings by the participants after interacting with the simulator for 20 iterations

## VIII. Conclusion

In this paper, we presented three different deception strategies from a “goal” directed perspective on an autonomous mobile robot simulator. We proposed an adaptive deception algorithm that can deceive humans in the short and long run using four different strategies. Our user study validated the hypothesis that the proposed adaptive algorithm combined with the deceptive strategies can deceive humans in long run. The user study also showed that the interaction with the deceptive robot was entertaining for the users. The experiment designed in this paper was just an example of a simple game in which participants predicted the true goal of the robot. It can also be applied in other interactive competitive games where the computer can deceive the user using the adaptive algorithm and available deceptive strategies.

Naturally, this research is not without limitations. We tested the adaptive algorithm only for $\lambda$=0.5 because the entropy analysis showed this to be the optimal value. Moreover, the experimental design was quite simple with only two targets involved. In the future, we will implement deceptive trajectories in a multi-target environment with targets placed randomly on the map. This will create a more realistic real time strategy game environment, which will give more insight into the advantages of this approach in interactive computer games.

## Acknowledgments

The authors thank Dr. Eugene Boman at PSU for suggesting the evaluation metrics (13) and (14) and reviewing the paper.